\documentclass[12pt]{article}
\usepackage{amsfonts,amsmath,amssymb}
\usepackage{graphicx}

\newfont{\twelvemsb}{msbm10 scaled\magstep1}
\newfont{\eightmsb}{msbm8}
\newfam\msbfam
\textfont\msbfam=\twelvemsb \scriptfont\msbfam=\eightmsb
\catcode`\@=11
\def\Bbb{\ifmmode\let\next\Bbb@\else
\def\next{\errmessage{Use \string\Bbb\space only in math mode}}\fi\next}
\def\Bbb@#1{{\fam\msbfam{{#1}}}}

\newcommand{\be}{\begin{equation}}
\newcommand{\ee}{\end{equation}}
\newcommand{\ba}{\begin{eqnarray}}
\newcommand{\ea}{\end{eqnarray}}

\begin{document}

\sloppy
\renewcommand{\thefootnote}{\fnsymbol{footnote}}
\newpage
\setcounter{page}{1} \vspace{0.7cm}
\begin{flushright}
November 2007
\end{flushright}
\vspace*{1cm}
\begin{center}
{\bf  Non-linear integral equations in ${\cal {N}}=4$ SYM
\footnote {Contribution to the proceedings of the conference
RAQIS07, 11-14 September 2007, Lapth Annecy, France (talk given by M.Rossi).}} \\
\vspace{1.8cm} {\large Diego Bombardelli, Davide Fioravanti 
and Marco Rossi
\footnote{E-mail:bombardelli@bo.infn.it, fioravanti@bo.infn.it, rossi@lapp.in2p3.fr}}\\
\vspace{.5cm} {\em Sezione INFN di Bologna, Dipartimento di Fisica, Universit\`a di Bologna, \\
Via Irnerio 46, Bologna, Italy} \\
\end{center}
\renewcommand{\thefootnote}{\arabic{footnote}}
\setcounter{footnote}{0}
\begin{abstract}
{\noindent We survey and discuss the applications of the non-linear integral equation in the framework of the
Bethe Ansatz type equations which are conjectured to give the eigenvalues of the dilatation operator in
${\cal {N}}=4$ SYM. Moreover, an original idea (different from that of \cite {FMQR}) to derive a non-linear integral
equation is briefly depicted in Section 4.}
\end{abstract}
\vspace{10cm} 

\newpage

\section{Introduction}

The AdS/CFT correspondence \cite {MWGKP} states the equivalence
between a string theory on the curved space-time
$\text{AdS}_5\times\text{S}^5$ and a conformal quantum field
theory on its boundary. In particular, type IIB superstring theory
should be dual to ${\cal N}=4$ Super Yang-Mills theory (SYM) in four dimensions:
energies of string states correspond to anomalous
dimensions of local gauge invariant operators of the quantum field
theory.

One of the most important recent development in this context was
the discovery of integrability in both planar field
theory \cite{MZ} and string theory \cite {BPR}. In a nutshell, integrable models appear
as spin chain type Bethe equations to be satisfied by 'rapidities' which
parametrise on the one side \cite {BDS,BS,BES} composite operators
and their anomalous dimensions in planar ${\cal N}=4$ SYM and 
on the other side \cite {BPR,KMMZ,AFS,BHL,BES} the corresponding dual objects in
string theory, i.e. states and their energies, respectively.

Thanks to this discovery, one could start to use the powerful
technique of the Bethe Ansatz in order to compute anomalous
dimensions of long operators, giving incredible boost towards a
proof of the AdS/CFT correspondence. However, the majority of the results
(with the exception, to our knowledge, of \cite {HLPS}) concerns the calculation 
of only the leading term of anomalous
dimensions of arrays of $L$ operators in the limit $L \rightarrow
\infty$. In order to study the (physically relevant) operators
with a finite number $L$ of components, in a series of papers 
\cite {FFGR1,FFGR2,FR} we proposed to interpret the Bethe
equations in terms of the non-linear integral equation (NLIE).
The NLIE \cite {FMQR} allows to write exact
expressions for the eigenvalues of the observables for arbitrary
values of the length of the chain ($L$ in this context) and the
number of Bethe roots. This allows to perform numerical
evaluations for such eigenvalues as well as, in some cases, to
give explicit analytic expressions for them, at least in some particular
limits.

In this contribution, we want to discuss our main results on the
application of the NLIE technique in the context of ${\cal N}=4$ SYM and
sketch at the end (Section 4) a new approach to and kind of NLIE.
We start (Section 2) from the so-called BDS model, which historically \cite {BDS}
was the first proposal for the many loops asymptotic (i.e. up to the order
${\lambda }^{L-1}$, $\lambda $ being the 't Hooft coupling) description of 
anomalous dimensions in the $SU(2)$ sector.
In this case, the NLIE and the exact expressions 
for the eigenvalues of the observables \cite {FFGR1} have the same structure
as in models studied in the past. This is a consequence of two facts.
First, the scattering matrix between magnons, which appears on the right hand side of the
Bethe equations, depends only on the difference of their rapidities (Bethe roots).
Second, for the state at exam the Bethe roots completely fill the real axis and allow only 
the presence of a finite number of holes. 
However, at least one of these properties fails when considering general states, or other 
sectors of ${\cal N}=4$ SYM or when higher loops corrections are improved by introducing the
so-called dressing factor \cite {BES}. 
In Section 3 we show how the NLIE and the corresponding expressions for the 
eigenvalues of the observables \cite {FR} in the $SU(2)$ sector assume a new and more complicated form
as a consequence of the failure of the first of previous properties, due to the presence of the dressing factor.
In order to cope with the intriguingly intricated structure of the Bethe equations and solutions appearing in 
${\cal N}=4$ SYM, in Section 4 we propose a new path to a 
NLIE substantially different from the idea of \cite {FMQR}. This procedure is effective when the magnon 
scattering matrix has a general dependence on the rapidities and the Bethe roots are concentrated 
on intervals of the real axis or complex lines.

\section{The NLIE for the BDS model}
\setcounter{equation}{0}

We start by considering the $SU(2)$ scalar sector of planar ${\cal N}=4$ 
Super Yang-Mills. In  a first proposal \cite {BDS}, composite operators (i.e. arrays of $L$ scalar fields)
were parametrised by rapidities $\{ u_k \}_{k=1,...,M}$ satisfying
the asymptotic \footnote {The term asymptotic means exactly
that this Ansatz is believed to give the exact loop expansion to
the anomalous dimension up to {\it wrapping corrections}, starting
at order $\lambda^L$.} Bethe-type equations
\begin{equation}
\label{BDS}
\left[\frac{X(u_j+\frac{i}2)}{X(u_j-\frac{i}2)}\right]^L=
\mathop{\prod^M_{k=1}}_{k\neq j} \frac{u_j-u_k+i}{u_j-u_k-i} \, ,
\end{equation}
where we introduced the function \be\label{ics} X(u)=\frac{u}{2}
\left( 1+\sqrt{1-\frac{\lambda}{4\pi^2 u^2}} \right)\, ,  \ee
$\lambda=Ng_{YM}^2=16 \pi ^2 g^2$ being the 't
Hooft coupling of planar theory ($N\rightarrow \infty,
g_{YM}\rightarrow 0$).
Anomalous dimensions (i.e. eigenvalues of the dilatation operator)
are given by
\begin{equation}
\Delta = L + \sum _{k=1}^M \left [\left (\frac {1}{2}-iu_k \right)
{\sqrt {1+\frac {4g^2 }{\left (\frac {1}{2}-iu_k \right)^2}}}+
\left (\frac {1}{2}+iu_k \right) {\sqrt {1+\frac {4g^2 }{\left
(\frac {1}{2}+iu_k \right)^2}}} -1 \right ]  \, . \label {Delta}
\end{equation}

In \cite {FFGR1} we wrote, following the standard route pioneered by
\cite {FMQR}, the NLIE describing the highest anomalous dimension
operator and its excitations. 
We now go briefly through its derivation (for details, see \cite {FFGR1}).
We first introduce two functions
\begin{equation}
\phi (u)= i \ln \frac {i+u}{i-u}=2 \arctan u \, , \, \, {\mbox
{Im}}u <1 \, , \quad \Phi (u)=i \ln \frac {X\left (\frac
{i}{2}+u\right )}{X\left (\frac {i}{2}-u\right )} \, , \, \,
{\mbox {Im}}u <1/2 \, , \label{fidef}
\end{equation}
which allows to define the counting function 
as \be\label{Zloops} Z(u)=L\, \Phi(u)-\sum
_{k=1}^M \phi (u-u_k)\, . \ee
We now consider a state with $M$ real roots $u_k$ and a number $H=L-2M$ of
real holes $x_h$. Using the fact that 
\be
e^{iZ(u_k)}=e^{iZ(x_h)}=(-1)^{L-M-1} \, , \label {condi}
\ee
the sum over all the
real roots of this state of a function $f$ analytic in a strip around the real axis can be written as \cite {FMQR}
\begin{eqnarray}
\sum _{k=1}^{M}f(u _k)&=&-\int _{-\infty}^{\infty}\frac{dv}{2\pi}
\, f'(v)\, Z(v)+
\label {prloops}\\
&+&\int _{-\infty}^{\infty }\frac{dv}{\pi}\, f'(v){\mbox { Im}}
\ln \left [1+(-1)^{L-M}e^{iZ(v+i0)}\right] - \sum
_{h=1}^{H}f(x _h)\, . \nonumber
\end{eqnarray}
This may be applied to the sum in the counting function
(\ref{Zloops}) bringing
\begin{eqnarray}
Z(u)&=&L\, \Phi(u)- \int _{-\infty}^{\infty}\frac
{dv}{2\pi} \, \phi'(u-v) \,
Z(v)+\nonumber \\
&+&\int _{-\infty}^{\infty }\frac {dv}{\pi}\, \phi '(u-v){\mbox
{ Im}}
\ln \left [1+(-1)^{L-M}e^{iZ(v+i0)}\right] + \label{prog} \\
&+ & \sum _{h=1}^H \phi (u-x_h) \, .\nonumber
\end{eqnarray}
It is convenient to introduce the usual 
synthetic notation \be L(u)={\mbox { Im}}\ln \left
[1+(-1)^{L-M}e^{iZ(u+i0)}\right]\, . \nonumber \ee
After $u$ Fourier transforming all the terms and moving the first
convolution to the l.h.s., we obtain
\begin{equation}\label{Zfourier}
\hat Z(k)=\hat F(k)+ 2 \hat G(k) \hat L(k)
+ \sum _{h=1}^H e^{-ikx_h} \hat H(k)\, ,
\end{equation}
where ($P$ is the principal value distribution) 
\be \hat F(k)=L \frac {\pi}{i} P\left ( \frac {1}{k} \right ) \frac {J_0 
\left(\frac{\sqrt{\lambda}}{2\pi}k\right)} {\cosh \frac {k}{2} }  \, , \quad \hat G(k)=\frac {1}{1+e^{|k|}} \, , \quad 
\hat H(k)=\frac {2\pi}{i} P\left ( \frac {1}{k} \right ) \hat G(k) \, .
 \label
{fourier} \ee
Inverting the Fourier transforms of (\ref{Zfourier}) leads to the NLIE valid for
this multi-loop Bethe equations (and for the aforementioned states) \ba
Z(u)&=&F(u)+ \sum _{h=1}^H H(u-x_h)+ \label{nlieloops}\\
&+&2 \int _{-\infty}^{\infty} dv\ G(u-v)\ {\mbox {Im}}\ln \left
[1+(-1)^{L-M}e^{iZ(v+i0)}\right]  \nonumber  \,. \ea
Let us observe that in this case the structure of this NLIE is quite the same
as in many other models. And, as usual, plugging the equation (\ref {nlieloops}) in relation (\ref {prloops}),
we can disentangle the bulk term (proportional to the size $L$) to the finite size corrections
in the expression for the eigenvalues of the observables:
\begin{gather}
\sum _{k=1}^{M}f(u _k)=-\int _{-\infty}^{\infty}\frac {dv}{2\pi}
f'(v) F(v) +  \label{fmultiloops}\\
+\int _{-\infty}^{\infty }\frac {dv}{\pi} f'(v)\int
_{-\infty}^{\infty} dw\ [\delta (v-w)-G(v-w)]{\mbox { Im}}\ln
\left [1+(-1)^{L-M}e^{iZ(w+i0)}
\right] - \notag \\
- \sum _{h=1}^H \left\{\int_{-\infty}^{\infty}\frac{dv}{2\pi}
f'(v) \, H (v-x_h) + f(x_h) \right\} \, .  \notag
\end{gather}
It is of interest to apply this formula to the conserved charges of the model, 
i.e.
\be
{\cal Q}_r=\sum _{k=1}^M q_r(u_k) \, , \quad q_r(u) =\int _{-\infty}^{\infty}\frac {dk}{2\pi} e^{iku} \
2^{r-1}\frac {(2\pi)^r}{({\sqrt {\lambda}})^{r-1}}\frac {1}{i^{r-2}}\frac {J_r\left (\frac {\sqrt {\lambda}}{2\pi}k \right )}{ke^{\frac {|k|}{2}}} \label {qrk} \, . 
\ee
In particular the anomalous dimension is given by $\Delta=L+2g^2 {\cal Q}_2$.
We get
\begin{eqnarray}
&&{\cal Q}_r=L \frac {i^{r+2}}{g^{r-1}}  \, \int
_{-\infty}^{\infty} dk \frac {J_{r-1}
( 2 g k) J_0( 2 g k)}{k (e^{|k|}+1)} - \int _{-\infty}^{\infty} \frac {dk}{2\pi} \frac
{i^{3+r}}{g^{r-1}}\frac {J_{r-1}(2gk)}{\cosh \frac {k}{2}} \hat
L(k) - \nonumber \\
&-& \sum _{h=1}^H \int _{-\infty}^{\infty} \frac {dk}{2k}
e^{-ikx_h} \frac {i^{2+r}}{g^{r-1}}\frac {J_{r-1}(2gk)}{\cosh
\frac {k}{2}} \, . \label{eqcharg1}
\end{eqnarray}
This relation is exact and allows the (numerical and analytical) study of the dependence of
the eigenvalues of the charges on the size of the system.
In particular, when $L\rightarrow \infty$, such eigenvalues behave as \cite {FFGR1}
\begin{eqnarray}
&&{\cal Q}_r=L \frac {i^{r+2}}{g^{r-1}}  \, \int
_{-\infty}^{\infty} dk \frac {J_{r-1}
( 2 g k) J_0( 2 g k)}{k (e^{|k|}+1)}- H \sum _{l=0}^{\infty} \frac {i^{2l+2-r}\pi ^{2l+r-1} g^{2l}}{l!(r+l-1)!}|E_{2l+r-2}| - \nonumber \\
&-& \frac {i^{r-1}}{g^{r-1}}\frac {J_{r-1}(2gi\pi)}{J_0(2gi\pi)} \frac {\pi}{12L} [1+(-1)^r] + o\left (\frac {1}{L} \right) \, ,
\end{eqnarray}
where $E_l$ are the Euler numbers.

\section{The NLIE in the $SU(2)$ sector with dressing factor}
\setcounter{equation}{0}

In order to match with results in string theory, the asymptotic Bethe Ansatz type
equations describing planar ${\cal N}=4$ SYM should contain --
with respect to the first proposals -- a universal dressing phase
\cite {AFS, BHL, BES}. Consequently, in the $SU(2)$ scalar sector
the BDS equations (\ref {BDS}) ought to be modified into
\begin{equation}
\label{BES} \left[
\frac{X(u_j+\frac{i}2)}{X(u_j-\frac{i}2)}\right ]^L=
\mathop{\prod^M_{k=1}}_{k\neq j} \frac{u_j-u_k+i}{u_j-u_k-i} \,
{\mbox {exp}}[2i\theta (u_j,u_k)] \, .
\end{equation}
On the other hand, the (renormalised) dimension corresponding to
the operator/solution $\{ u_k \}_{k=1,...,M}$ of (\ref {BES})
is still given by (\ref {Delta}).
The complete dressing phase has been
conjectured to be 
\cite{BES}
\begin{equation}
\theta (u_j,u_k)= \sum _{r=2}^{\infty} \sum _{\nu
=0}^{\infty} \beta _{r,r+1+2\nu}(g)
[q_r(u_j)q_{r+1+2\nu}(u_k)-q_r(u_k)q_{r+1+2\nu}(u_j)] \, .
\label{theta-beta}
\end{equation}
In order to fix the notations in (\ref
{theta-beta}), we remind that $q_r(u)$ is the magnon $r$-th
charge, which is given by formula (\ref {qrk})
and the functions $\beta _{r,r+1+2\nu}(g)$ are meromorphic
functions of $g$, introduced and studied in \cite {BES}. In that
paper their weak coupling expansion was proposed as
\begin{equation}
\beta _{r,r+1+2\nu}(g)=\sum _{\mu=\nu}^{\infty}
{g}^{2r+2\nu+2\mu}\beta _{r,r+1+2\nu}^{(r+\nu+\mu)} \, ,
\label{betaexp}
\end{equation}
the coefficients $\beta _{r,r+1+2\nu}^{(r+\nu+\mu)}$ being
\begin{equation}
\beta _{r,r+1+2\nu}^{(r+\nu+\mu)}=2(-1)^{r+\mu+1}\frac
{(r-1)(r+2\nu)}{2\mu+1} \left ( \begin{array}  {c} 2\mu +1 \\
\mu-r-\nu+1 \end{array} \right) \left ( \begin{array} {c} 2\mu +1
\\\mu - \nu \end{array} \right ) \zeta (2\mu+1) \, .
\end{equation}
In \cite {FR} we wrote a NLIE equivalent to the Bethe equations (\ref {BES})
for the states\footnote {As in the BDS model, these states are 
the highest anomalous dimension state and its excitations.} characterised by $M$ real Bethe
roots, $u_k$, and $H=L-2M$ real holes, $x_h$. We now go briefly through its derivation.
We start by defining the counting function as
\be\label{Zdef} Z(u)=L\, \Phi(u)-\sum
_{k=1}^M \phi (u-u_k)+2\sum _{k=1}^M\theta (u,u_k)\, . \ee
Then, as a consequence of (\ref {condi}), it is simple to express a sum
on the Bethe roots for a function $f(u)$ as 
\begin{equation}
\sum ^M_{k=1} f(u_k)=- \int _{-\infty}^{\infty} \frac {dv}{2\pi} \
\frac {d}{dv}f(v) \ [Z(v)-2L(v)]-\sum _{h=1}^H f(x_h) \, . \label
{fsum}
\end{equation}
In particular, we will be
interested in the eigenvalues of the conserved charges
\begin{equation}
{\cal Q}_r=\sum ^M_{k=1} q_r(u_k)=- \int _{-\infty}^{\infty} \frac
{dv}{2\pi} \ \frac {d}{dv}q_r(v) \ [Z(v)-2L(v)] -\sum _{h=1}^H q_r
(x_h) \, . \label {qrsum}
\end{equation}
Applying (\ref {fsum}) to (\ref {Zdef}) we get
\begin{eqnarray}
Z(u)&=&L\Phi (u) - \int _{-\infty}^{\infty} \frac {dv}{2\pi} \frac
{2}{(u-v)^2+1}[Z(v)-2L(v)] + \sum _{h=1}^H \phi (u-x_h)-  \nonumber \\
&-& 2 \int _{-\infty}^{\infty} \frac {dv}{2\pi} \ \frac {d}{dv}
\theta (u,v) \ [Z(v)-2L(v)] -2\sum _{h=1}^H \theta (u,x_h)\, .
\nonumber
\end{eqnarray}
Going now to the Fourier space, after grouping the terms 
containing $\hat Z(k)$, the following equation shall hold
\begin{eqnarray}
\hat Z(k)&=& L \frac {2\pi}{i} P\left (\frac {1}{k}\right ) \frac
{J_0(2gk)}{2\cosh \frac {k}{2} }  + \frac {2}{1+e^{|k|}} \hat
L(k) + \sum _{h=1}^H e^{-ikx_h} \frac {2\pi}{i}  P\left (\frac {1}{k}\right ) \frac {1}{1+e^{|k|}} +\nonumber  \\
&+&\frac {1}{\cosh \frac {k}{2}} \sum _{r=2}^{\infty}\sum
_{\nu=0}^{\infty} {\beta}_{r,r+1+2\nu}(g) \Bigl [\frac
{2\pi }{g^{r-1}}\frac {1}{i^{r-2}}\frac
{J_{r-1}(2gk)}{k}{\cal Q}_{r+1+2\nu}- \nonumber \\
&-& \frac {2\pi }{g^{r+2\nu}}\frac {1}{i^{r+2\nu-1}}\frac
{J_{r+2\nu}(2gk)}{k}{\cal Q}_r\Bigr ] \, , \label {Zeq2}
\end{eqnarray}
where we have introduced the conserved charges (\ref {qrsum}) and the explicit form (\ref {qrk}) of the Fourier transform
of the charge densities, $\hat q_r(k)$, has been used. A very crucial difference
of this non-linear integral equation from the others in the
literature may be stated in the presence of $Z(u)$ in infinite
many places, i.e. all the charges ${\cal Q}_r$ (\ref{qrsum}).

Concerning the charges, we can write for them a system of
equations. In fact, we first rewrite the
expressions (\ref{qrsum}) in terms of Fourier transforms
\begin{eqnarray}
{\cal Q}_s&=&-\int _{-\infty}^{\infty} \frac {dk}{4\pi^2} \hat
{q^{\prime}_s} (-k) [\hat Z(k) -2 \hat L(k) ] -\sum _{h=1}^H q_s(x_h)= \nonumber \\
&& = \int _{-\infty}^{\infty} \frac {dk}{2\pi} \frac
{i^{3+s}}{g^{s-1}}\frac {J_{s-1}(2gk)}{e^{ \frac {|k|}{2}}} [ \hat
Z(k) - 2 \hat L(k) ] -\sum _{h=1}^H q_s(x_h) \, .
\end{eqnarray}
Then we insert relation (\ref {Zeq2}) for $\hat Z(k)$ into this expression to
obtain, 
\begin{eqnarray}
&&{\cal Q}_s=\frac {i^{s+2}}{g^{s-1}} \Bigl [ L \, \int
_{-\infty}^{\infty} dk \frac {J_{s-1}
( 2 g k) J_0( 2 g k)}{k (e^{|k|}+1)} + \nonumber \\
&+& 2  \sum _{r=2}^{\infty} \sum _{\nu =0}^{\infty}
{\beta}_{r,r+1+2\nu}(g) (-1)^{1+\nu} \int _{-\infty}^{\infty} dk \frac
{J_{s-1} ( 2 g k) J_{r-1}( 2 g k)}{k (e^{|k|}+1)}  \frac {
g^{1-r}}{ i^{r+2\nu-1}} {\cal
Q}_{r+2\nu+1}  \nonumber \\
&+&  2 \sum _{r=2}^{\infty} \sum _{\nu =0}^{\infty}
{\beta}_{r,r+1+2\nu}(g) (-1)^{1+\nu} \int _{-\infty}^{\infty} dk \frac
{J_{s-1} ( 2 g k) J_{r+2\nu}( 2 g k)}{k (e^{|k|}+1)} \frac {
g^{-r-2\nu}}{ i^{r-2}} {\cal
Q}_{r} \Bigr ] - \nonumber \\
&-& \int _{-\infty}^{\infty} \frac {dk}{2\pi} \frac
{i^{3+s}}{g^{s-1}}\frac {J_{s-1}(2gk)}{\cosh \frac {k}{2}} \hat
L(k) -\sum _{h=1}^H \int _{-\infty}^{\infty} \frac {dk}{2k}
e^{-ikx_h} \frac {i^{2+s}}{g^{s-1}}\frac {J_{s-1}(2gk)}{\cosh
\frac {k}{2}} \, . \label{eqcharg2}
\end{eqnarray}
This relation is exact and, at least in principle, may be
efficient in the analysis of the conserved charges, though now
$Z(u)$ appears.

\section{A new approach to a NLIE}
\setcounter{equation}{0}

In almost all the cases considered up to now, the NLIE was written
for counting functions defined as \be Z(u)=\Phi (u)-\sum
_{k=1}^{M} \phi (u-u_k) \, , \ee and when the Bethe roots
distribute on the real axis, allowing the presence of only a
finite number of holes and possibly complex roots. Even if this case is relevant for the
study of the fundamental state and the first excitations 
of many models, it does not cover many of the Bethe Ansatz systems
proposed in the context of ${\cal N}=4$ SYM.

For this reason we want to write the NLIE (and the expression for the 
eigenvalues of the observables in terms of its solution) for the more general
case in which the counting function is defined as \be Z(u)=\Phi
(u)-\sum _{k=1}^{M} \phi (u,u_k) \, , \label {Z2} \ee (i.e. the function $\phi
(x,y) $ does not depend only on the difference $x-y$: this happens, for instance, 
when the dressing factor is present). We suppose
also that the $M$ Bethe roots are
concentrated in an interval $[A,B]$ of the real axis \footnote {The case in which 
the Bethe roots are concentrated on a finite number of intervals on the real axis follows straightforwardly
from the results of this Section. Moreover, even the case when the roots lie on complex lines
can be treated as follows.} and that holes are present only outside this interval. This second
property is peculiar, for instance, of some states in the $sl(2)$ sector of
${\cal N}=4$ SYM.

On this state we consider a sum over the Bethe roots $\{ u_k \}_{k=1,...,M}$ 
of a function (observable) $O(u)$ analytic in a strip around the real axis. If the condition
$e^{iZ(u_k)}=-1$ holds, this sum can be written as
\begin{eqnarray}
&& 2\pi i \sum _{k=1}^{M}O(u _k)= \lim _{\epsilon \rightarrow 0^+} \left [ \int _{A}^{B}du  O(u-i\epsilon)\frac {e^{iZ(u-i\epsilon)}iZ^{'}(u-i\epsilon)}{1+e^{iZ(u-i\epsilon)}}+ \right. \label {equ2} \\
&+& 
\left. \int _{B}^{A}du  O(u+i\epsilon)\frac {e^{iZ(u+i\epsilon)}iZ^{'}(u+i\epsilon)}{1+e^{iZ(u+i\epsilon)}} \right ] \, .
\nonumber
\end{eqnarray}
Supposing $Z^{'}(u)<0$, we can rearrange this
expression as follows,
\begin{eqnarray}
\sum _{k=1}^{M}O(u _k)&=&-\int _{A}^{B}\frac {dv}{2\pi}
O(v)Z^{'}(v)+\int _{A}^{B}\frac {dv}{\pi} O(v)\frac {d}{dv}{\mbox {Im}}\ln \left
[1+e^{iZ(v-i0)}\right] = \nonumber \\
&=& - \frac {1}{2\pi}\left [O(B)Z(B)-O(A)Z(A) \right] +\nonumber \\
&+&\frac {1}{\pi} \left \{ O(B)
{\mbox {Im}}\ln \left [1+e^{iZ(B)}\right]- O(A)
{\mbox {Im}}\ln \left [1+e^{iZ(A)}\right] \right \} + \nonumber \\
&+& \int _{A}^{B}\frac {dv}{2\pi} O^{'}(v) Z(v)-\nonumber \\
&-&2\int _{A}^{B}\frac {dv}{2\pi} O^{'}(v){\mbox
{Im}}\ln \left [1+e^{iZ(v-i0)}\right] \, . \label {prop2}
\end{eqnarray}
In brief, what we are doing is to evaluate a sum on the Bethe roots by integrating just on the interval containing them.
Therefore, this method is alternative and complementary to the idea proposed in the first of \cite {FMQR} which consists in first integrating on all the real axis and then subtracting the contributions coming from the real holes.

We now apply (\ref {prop2}) to the sum over the Bethe roots appearing
in the definition (\ref {Z2}). We get the following equation
\begin{eqnarray}
Z(u)&=&f(u) - \int _{A}^{B}\frac {dv}{2\pi} \, \frac {d}{dv} \phi (u,v)\, Z(v)+\nonumber \\
&+&2\int _{A}^{B}\frac {dv}{2\pi} \, \frac {d}{dv} \phi (u,v)\, {\mbox
{Im}}\ln \left [1+e^{iZ(v-i0)}\right] \, , \label{2eq1}
\end{eqnarray}
where 
\begin{eqnarray}
f(u)&=&\Phi (u)+\frac {1}{2\pi}\left [ \phi (u,B)Z(B)-\phi (u,A)Z(A) \right] -\nonumber \\
&-&\frac {1}{\pi} \left \{ \phi (u,B)
{\mbox {Im}}\ln \left [1+e^{iZ(B)}\right]- \phi (u,A)
{\mbox {Im}}\ln \left [1+e^{iZ(A)}\right] \right \}  \, . 
\end{eqnarray}
We can now write a NLIE
for the counting function by inserting in an iterative way (\ref
{2eq1}) for $Z$ in the right hand side of the same equation. Using the notation
\begin{equation}
(\varphi \star f) (u) = \int _{A}^{B} dv \, \varphi (u,v)f(v) \, ,
\end{equation}
eventually we get the NLIE in the form
\begin{equation}
Z(u)= F (u) + 2 (G \star L) (u) \, , \label {nlin}
\end{equation}
where
\begin{equation}
F (u)= f(u)+\sum _{k=1}^{\infty} (-1)^k ((\varphi ^{\star k}) \star f )(u)
\, , \quad G(u,v)=  \varphi (u,v)+ \sum _{k=2}^{\infty} (-1)^{k-1} \, (\varphi
^{\star k} )(u,v) \, . \label {FGdef}
\end{equation}
We used the simplified notations
\begin{equation}
L(u)={\mbox {Im}}\ln \left [1+e^{iZ(u+i0)}\right] \, , \quad 
\varphi (u,v) = \frac {1}{2\pi} \frac {d}{dv} \phi (u,v)  \, . 
\end{equation}
More explicitly,
\begin{eqnarray}
F(u)&=& f(u)+\sum _{k=1}^{\infty}(-1)^k \int _{A}^{B}dv_1 \, \varphi (u,v_1)  \int _{A}^{B}dv_2 \, \varphi (v_1,v_2) \ldots \nonumber \\
&\ldots &
\int _{A}^{B}dv_k \, \varphi (v_{k-1},v_k) f(v_k) \, , \label {Fdef} \\
G(u,v)&=&\varphi (u,v)+\sum _{k=1}^{\infty} (-1)^k \int _{A}^{B}dv_0 \, \varphi (u,v_0)\int _{A}^{B}dv_1\, \varphi (v_0,v_1) \ldots \nonumber \\
&\ldots & \int _{A}^{B}dv_{k-1} \, \varphi (v_{k-2}-v_{k-1}) \, 
\varphi (v_{k-1}-v) \, . \label {Gdef}
\end{eqnarray}
Eventually, inserting (\ref {nlin}) in (\ref {prop2}) we get an expression for the eigenvalues of an observable as
\begin{eqnarray}
\sum _{k=1}^M O(x_k)&=&
- \frac {1}{2\pi}\left [O(B)Z(B)-O(A)Z(A) \right] +\nonumber \\
&+&\frac {1}{\pi} \left \{ O(B)
{\mbox {Im}}\ln \left [1+e^{iZ(B)}\right]- O(A)
{\mbox {Im}}\ln \left [1+e^{iZ(A)}\right] \right \} + \nonumber \\
&+& \int _{A}^{B}\frac {dv}{2\pi}   O^{'}(v) F(v)+\label {Oexp} \\
&+&2\int _{A}^{B}\frac {dv}{2\pi}  O^{'}(v)\int _{A}^{B} dw [G(v,w)-\delta (v-w)]{\mbox
{Im}}\ln \left [1+e^{iZ(w-i0)}\right] \, . \nonumber 
\end{eqnarray}
We remark that all the already known NLIEs can be reproduced in this way, without Fourier transforming.
Moreover, formul{\ae} (\ref {nlin}, \ref {FGdef}) and (\ref {Oexp}) can be used in order to write, respectively, the NLIE and the eigenvalues of the observables on states appearing in models relevant for ${\cal N}=4$ SYM.
It would be of interest to apply these techniques, for instance, to the widely studied \cite {ES,BES} $sl(2)$ sector of the theory. 

\section{Summary}

We have reported on our project - still in progress - of writing NLIEs for the Bethe Ansatz type equations relevant for the 
determination of anomalous dimensions of operators in ${\cal N}=4$ SYM (and, correspondingly
energies of strings). The NLIE describing the highest anomalous dimension operator of the BDS model and its excitations - treated in Section 2 - does not differ in form from the NLIEs studied up to now.
However, the Bethe equations arising in the context of ${\cal N}=4$ SYM have in general intrinsic complications - namely, magnon scattering matrix which does not depend only on the 
difference of the rapidities and states described by roots which can condense on lines in the complex plane - which at first sight seem to prevent even the possibility of writing a NLIE.
We have shown how to circumvent this problem in some cases. First we studied in Section 3 the $SU(2)$ sector with dressing factor and 
showed how to write a new type of NLIE which depends explicitly on the eigenvalues of the conserved charges. Then, in Section 4,
we sketched  a general formalism which allows to treat magnon scattering matrices with general dependence on the rapidities and states with roots on intervals on the real line (or even complex lines). We plan to give in forthcoming publications explicit 
applications of this new formalism, for instance in the $sl(2)$ sector of ${\cal N}=4$ SYM.

\end{document}